\documentstyle{mn}

\input psfig

\begin{document}

\title[Two new $z>4$ radio-loud Quasars]{Discovery of radio-loud quasars with z=4.72 and z=4.01}

\author[I. M. Hook and R. G. McMahon]
{\parbox[]{6.in} 
{Isobel M. Hook$^1$ \thanks{Present address: European Southern
Observatory, Karl Schwarzschild Stra\ss{e} 2, D-85748 Garching b. M\"{u}nchen, Germany} and Richard G. McMahon$^2$}  \\
       $^1$ U.C. Berkeley Astronomy Dept, Berkeley, CA94720, U.S.A \\
       $^2$ Institute of Astronomy, Madingley Road, Cambridge CB3 0HA,
UK \\
       {\rm email: ihook@eso.org, rgm@ast.cam.ac.uk}}

\date{submitted: 25 April 1997; version \today}

\maketitle

%\date{version \today}

\begin{abstract}
  We report the discovery of two radio-loud quasars with redshifts
  greater than four; GB1428+4217 with $z=4.72$ and GB1713+2148 with
  $z=4.01$.  This doubles the number of published radio-selected
  quasars with $z>4$, bringing the total to 4.
  GB1428+4217 is the third most distant quasar known and the highest
  redshift radio and X-ray source currently known. It has a radio flux
  density at 5GHz of $259\pm 31$ mJy and an optical magnitude of $\rm
  R\sim 20.9$. The rest frame absolute UV magnitude, $\rm
  M_\nu(1450\AA)$, is $-$26.7 similar to that of the archetypal
  radio-selected quasar 3C273 ($z=0.158$; $\rm M_\nu(1450\AA)=-26.4)$.
GB1428+4217 has a tentative detection in ROSAT PSPC observations,
which has been confirmed by more recent ROSAT observations, described
in a companion paper by Fabian et al.  Both quasars were discovered
during the CCD imaging phase of an investigation into the evolution of
the space density of radio-loud quasars at high redshift. Combined
with our earlier survey results these objects give a lower limit on
the space density of quasars with radio power $\rm P_{5GHz}> \rm 5.8
\times 10^{26} W Hz^{-1} sr^{-1}$ between
$z=4$ and $z=5$ of $\rm 1.4 \pm 0.9\times 10^{-10} Mpc^{-3}$.  This
can be compared to $\rm 2.9 \pm 0.2\times 10^{-10} Mpc^{-3}$ at $z=2$ from
Dunlop \& Peacock (1990) for flat-spectrum sources of the same
luminosity.

\end{abstract} 

\maketitle

\section{Introduction}

In two recent papers we have described results from the first phase of
a survey for high-redshift, radio-loud quasars (Hook et al 1995,
1996). Our approach involves the optical identification of flat
spectrum radio sources and the spectroscopic follow-up of the red
stellar identifications. This approach exploits the observation that
quasars at high redshift have redder optical colours than their
low-redshift counterparts due to absorption by intervening
HI (see figure 1 in Hook {\it et al.}  1995). 

Our aim is to study the evolution of the quasar population
at high redshift, using a well-defined, statistically complete
sample. Whilst radio-loud quasars are only a small subset
of the quasar population, the selection of radio-loud quasars is less
prone to selection effects than optical samples, since the spectral
energy distribution of radio sources is smooth and radio emission is
unaffected by either intrinsic or extrinsic absorption due to
dust. See Fall \& Pei (1995) for a discussion of how dust within
intervening galaxies may affect the observed evolution in optically
selected samples of quasars. 
Our work complements the numerous investigations into the redshift
evolution of the space density of radio quiet quasars (eg.  Hall et
al., 1996; Hawkins \& Veron, 1996; Irwin, McMahon \& Hazard, 1991;
Kennefick et al., 1996; Schmidt, Schneider \& Gunn, 1995; Warren,
Hewett \& Osmer, 1995).

Our earlier work was restricted to sources that were detected on APM
scans of POSS-I plates and hence was incomplete for optically fainter
quasars.  Since it is our aim to determine the evolution of the space
density of high-redshift quasars independent of optical selection
effects, we have now begun CCD imaging of sources that were not
detected on the POSS-I plates.

Throughout this paper we have assumed cosmological constants of
$H_0={\rm 50\,km\,s^{-1}\,Mpc^{-1}}$ and $q_0=0.5$.

\section{The Radio Sample and the Optical Identification procedure}

The parent radio sample used in this study consists of $\sim$1600
flat-spectrum ($\rm \alpha^{5GHz}_{1.4GHz} \ge-0.5$, $\rm S\propto
\nu^{\alpha}$) radio sources with $\rm S_{5GHz}>0.2Jy$ studied with the
VLA by Patnaik {\it et al.} (1992).  Initial optical identification of
the radio sources was carried out using the Automated Plate
Measurement (APM) Facility at Cambridge, U.K. Further details of this
program and the procedures can be found in Hook et al (1995,1996).

For the second phase of the survey, we have concentrated on the radio
sources with no APM/POSS-I identification within a search radius of
$3.0''$.  The area studied lies in the region with $\rm 9^h 50' <
\alpha(B1950) < 17^h 45'$, $\rm 20^{\circ} < \delta(B1950) < 75^{\circ}$,
($|b| > 20^{\circ}$ for $0^h$ to $12^h$ and $|b| > 30^{\circ}$ for
$12^h$ to $24^h$), an area of 1.27sr containing 526 sources.  There
were 82 blank fields with $\rm S_{5GHz}>0.2Jy$ in this region, after
our earlier work. Of these, 16 had redshifts from other sources prior
to the start of this project (NED\footnote{The NASA/IPAC
Extragalactic Database (NED) is operated by the Jet Propulsion
Laboratory, California Institute of Technology, under contract with
the National Aeronautics and Space Administration.}, Vermeulen et al
1996, Vermeulen, private communication).

An unbiased sub-sample of 38 of the remaining 66 sources were imaged
during one week (29 May to 4 June 1995) at the Prime Focus of 2.5m
Isaac Newton Telescope, La Palma.  We used the $1024 \times 1024$
thinned TEK CCD with a projected pixel size of $0.59''$. Mould B and R
filters (The `Kitt-Peak' set at La Palma) were used with 300s
exposures in R and 600s exposures in B, and typically reaching
limiting magnitudes of $\rm R=23$mag, $\rm B=24$mag.

Photometric calibration was carried out using observations of the
photometric standard star fields from Landolt (1992). The data were
reduced using standard procedures with the IRAF\footnote{IRAF is
distributed by the National Optical Astronomy Observatories, which is
operated by the Association of Universities for Research in Astronomy,
Inc. (AURA) under cooperative agreement with the National Science
Foundation.} software environment including photometry using the
DAOPHOT package (Stetson, 1990). 

\begin{figure} 
\centerline{\psfig{figure=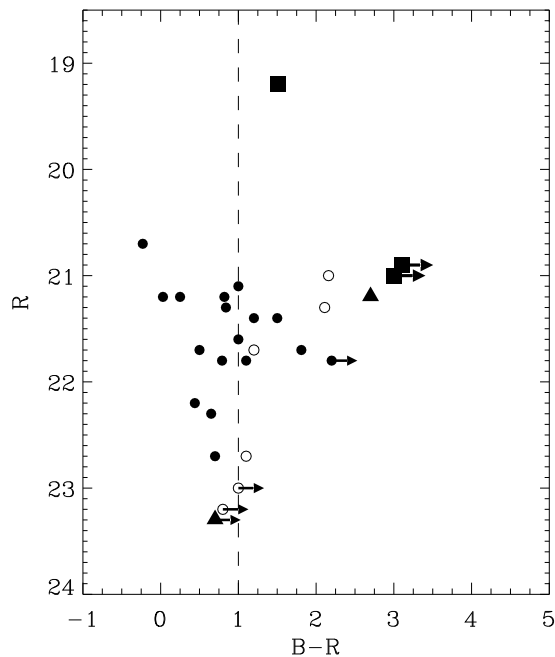,height=3.5in,bbllx=20pt,bblly=288pt,bburx=247pt,bbury=555pt}}
\caption{Colour-magnitude diagram from INT CCD identifications for the $\rm S_{5GHz}>0.2$Jy
flat-spectrum radio sample. Filled circles show unresolved sources,
open circles represent galaxies. The squares show
spectroscopically-confirmed $z>3$ quasars. The dashed line indicates
the boundary used for the spectroscopic subsample.  The filled
triangles show the 2 objects of in that sample for which a spectrum
has yet to be obtained (see text).}
\label{colmag}
\end{figure}

After this INT imaging, 10 sources are still blank in the R band,
i.e. they had no optical counterpart within a radius of $3''$ (K band
imaging of these sources is being obtained). A colour magnitude
diagram for the 28 objects detected in R is shown in
Figure~\ref{colmag}.

Thus after the INT imaging the fraction of sources observed is $38/66=
0.58$. To estimate in a fair manner the effective area observed, we
include the same fraction (9 out of 16) of the sources with previously
known redshifts. The effective area is then $0.58 \times
1.27sr=0.73sr$. Scaling from the original 526 radio sources in 1.27sr,
there are 302 radio sources in this sub-area of which all but 10 (3\%)
are now identified.

\section{Spectroscopic observations and results}

We have selected for spectroscopic followup a complete sample of the
12 stellar identifications with $\rm B-R \ge 1.0$.  Redshifts became
available for 3 of the 12 selected objects prior to the spectroscopic
phase of this study (Vermeulen, private communication) including the
$z=3.82$ quasar GB1239+3736 (Vermeulen et al 1996).
7 of the other 9 were observed at either the Shane 3m
telescope at Lick observatory or the 4.2m William Herschel Telescope
(WHT), La Palma. The Lick spectra were obtained using the KAST double
spectrograph at lowest resolution covering the range $5000-10500$\AA.
The WHT spectra were obtained using the red arm of the ISIS
double spectrograph.  A 158 l/mm grating and a thinned TEK CCD with
1024x1024 $\rm 24\mu m$ pixels was used, giving a useful wavelength
range of 5970-8940\AA\ at 2.9\AA\ per pixel.

\begin{figure*}
\centerline{
\psfig{figure=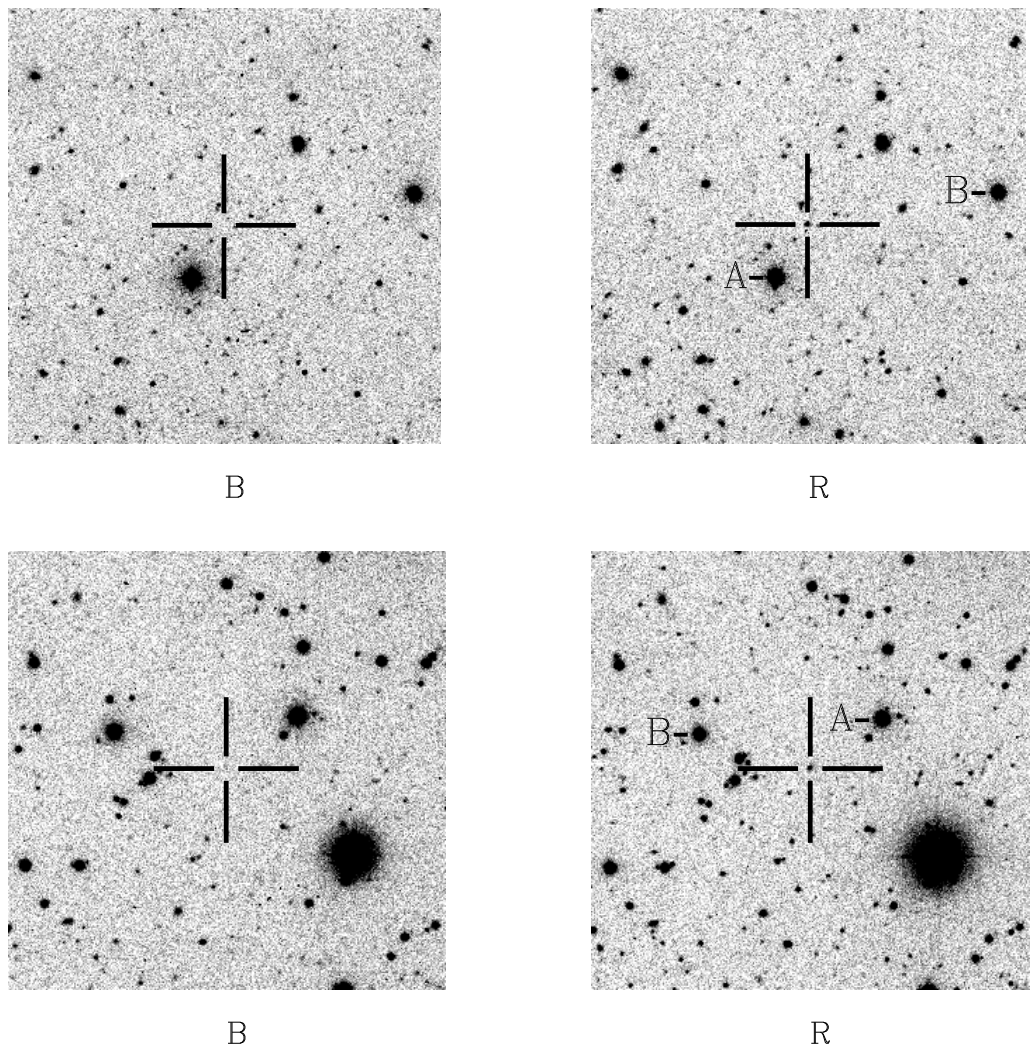,height=5.5in,bbllx=17pt,bblly=43pt,bburx=322pt,bbury=356pt}}
\caption{B and R band images, obtained at INT, of the two $z>4$
quasars. The upper panels show GB1428+4217 and the lower panels show
GB1713+2148. The central cross is 1 arcmin across. North is at the
top, West to the right. The offsets ($\Delta\alpha$, $\Delta\delta$)
in arcseconds from the stars marked A and B to the quasars are as
follows (positive $\Delta\alpha$ represents a move East from the
offset star to the quasar, and positive $\Delta\delta$ represents a
move North): GB1428+4217: from A to quasar ($-13.6$, $22.3$), from B to
quasar (80.7 $-12.7$); GB1713+2148: from A to quasar (30.2, $-19.9$)
from B to quasar ($-46.0$, $-14.6$).}
\label{images}
\end{figure*}

Of the 7 objects observed, 1 still has an inconclusive spectrum.
Another (GB1712+493) has a recently-published redshift of 1.552
(Falco, Kochaneck \& Mu\~noz 1997), consistent with the single
emission line seen in our spectrum being identified with MgII.  The
spectrum of one other object also shows a single broad emission line
(probably also MgII) which we can rule out as being Ly-$\alpha$ at
$z>4$ by the lack of other strong emission lines or Ly-$\alpha$ forest
absorption.  Here we give details for two objects, GB1428+4217 and
GB1713+2148, which were identified as high-redshift quasars.  The INT
B and R band images of the 2 new $z>4$ quasars are shown in
Figure~\ref{images}.

\begin{figure} 
\psfig{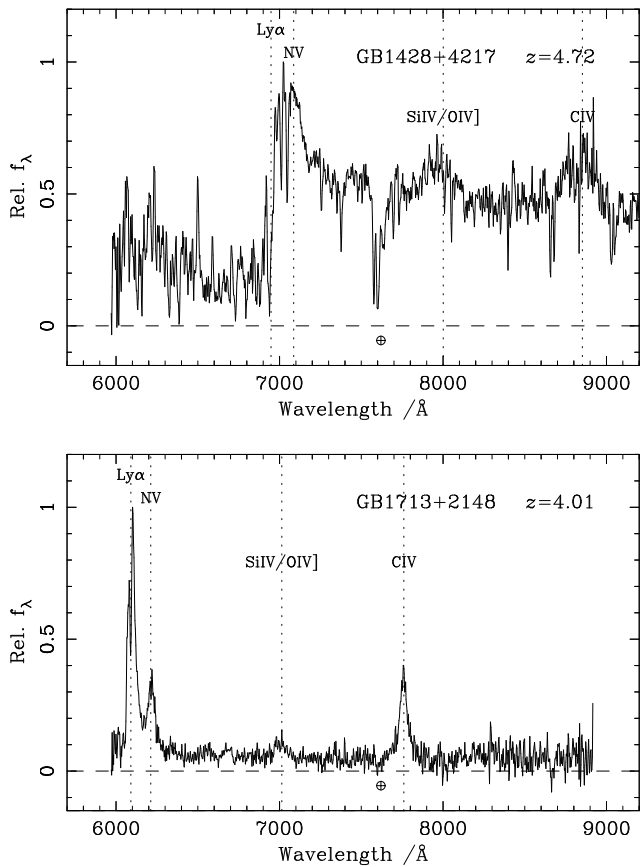}
\caption{Optical spectra of the two $z>4$ quasars.}
\label{spectra}
\end{figure}

\paragraph*{\bf GB1428+4217 ({\it z}=4.715$\pm$0.010)} GB1428+4217 was first
observed on 1996 16th July on the 4.3m WHT and then again on 1996 19th
July at a redder wavelength grating setting in order to detect the CIV
emission line.  Broad emission lines of Ly$\alpha$, NV, SiIV/OIV] and
CIV are visible in the spectrum (Fig.~\ref{spectra}). The observed
centroids and and equivalent widths are given in Table~1.

The Ly$\alpha$ line is visibly asymmetric due to absorption and it is
therefore impractical to measure the line centroid. If one assumes
that the absorption is due to intervening HI, the blue edge of the
line provides a lower limit to the redshift. The peak of the blue edge
occurs at 6974\AA. The correction for instrumental resolution effects
is $\rm \sim3\AA$, so assuming a value of 6971\AA\ for the Ly-$\alpha$
edge gives a redshift of 4.734. It is quite common to observe velocity
shifts between the emission lines in quasars and this redshift
difference corresponds to a blue shift of 1420 km/sec of CIV with
respect to Ly-$\alpha$. This is significantly higher that the mean
shift of 138 km/sec derived by Tytler and Fan (1992) in their analysis
of a sample of 160 quasars.  We take the redshift of GB1428+4217 to be
the mean of the three estimates.

On the night of 19th July, both narrow and wide slit (7 arc second)
spectrophotometric observations of GB1428+4217 were obtained.  The
flux standard G138-31 (1625+0919; Filippenko \& Greenstein, 1984) was
observed both prior to and immediately after the observations of
GB1428+4217. These observations give a continuum flux at a rest
wavelength of 1500\AA\ ($\lambda_{obs}$=8550\AA) of $\rm 6.3 \times
10^{-28} ergs~cm^{-2} s^{-1} Hz^{-1} $.  The corresponding continuum
apparent magnitude and absolute magnitude on the AB system (Oke \&
Gunn 1983) are
19.4 and $-$26.7 respectively. For comparison, from HST observations
of the archetypal radio-selected quasar 3C273 ($z=0.158$) (Bahcall et
al 1991) we find the continuum luminosity to be $\rm
M_\nu(1450)=-26.4$.

\paragraph*{{\bf GB1713+2148}~({\it z}=4.011$\pm$0.005)}
The discovery spectrum of GB1713+2148 was obtained at Lick-3m on UT
1996 May 23.  It was reobserved at the WHT on 1996 16th July for 1800s
to obtain a higher signal-to-noise spectrum, shown in
Figure~\ref{spectra}. Broad emission lines of Ly$\alpha$, NV,
SiIV/OIV] and CIV are visible in the spectrum. Again the Ly$\alpha$
line has visible absorption.

\begin{table*}
\label{twoqsos}
\caption{ Properties of the new $z > 4$ quasars. Redshifts were measured as described in the text,
assuming rest-frame wavelengths for the emission lines as follows:
Ly$\alpha$:1215.7, CIV: 1549.1, NV:1240.1, SiIV/OIV]:1397.8 (see
Tytler \& Fan 1992).  The radio fluxes are from Gregory and Condon
(1991).  The R and B magnitudes are accurate to 0.07mag (rms) or
better.  $*$=Observed equivalent width measured for the Ly$\alpha$-NV combination, $p$=
wavelength measured from the peak rather than the centroid of line.}
\begin{tabular}{lcccrrcccccc}
Name 
&\multicolumn{1}{c}{Optical position}
& R 
& $\rm B-R$ 
&\multicolumn{1}{c}{ $\rm S_{5GHz}$} 
& Feature 
&$\lambda_{obs}$ 
& $\rm W_\lambda$ 
&\multicolumn{1}{c}{$z$} 
& \multicolumn{1}{c}{$<z>$}
\cr 

& \multicolumn{1}{c} {$\alpha$ (B1950) $\delta$} 
& mag 
&mag
&\multicolumn{1}{c}{mJy} 
& 
&\AA 
&\AA 
& 
& 
\cr 
& & & & & & & & 
\cr 
GB1428+4217
& 14\ 28 26.74~~+42\ 17\ 52.6 
& 20.9 
& $>$3.40
& $259\pm 31$
&Ly$\alpha$ 
&$6971^p$
&$98^*$ 
&4.734 
\cr 
& & & & &NV & & &
\cr 
& & & & &SiIV/OIV] &7972\phantom{$^p$}&48\phantom{$^*$}
&4.707 & 
\cr 
& & & & &CIV &8841\phantom{$^p$}&74\phantom{$^*$}
&4.703 &  
\cr
&&&&&&&&&4.715$\pm0.010$ 
\cr
\cr
GB1713+2148 &17\ 13\ 13.66~~+21\ 48\ 51.4 & 21.0 & $>$2.90&
     $327\pm 44$ &Ly$\alpha$ &$6098^p$&$536^*$& 4.016 &
\cr 
& & & &
     &NV &$6216^p$& &4.013& 
\cr 
& & & & &SiIV/OIV] &6999\phantom{$^p$}&94\phantom{$^*$}&4.008 & 
\cr & & & & &CIV
     &7758\phantom{$^p$}&640\phantom{$^*$}&4.008 & 
\cr
&&&&&&&&&4.011$\pm0.005$ 
\cr
\end{tabular}\newline
\end{table*}

\begin{figure}
\centerline{\psfig{figure=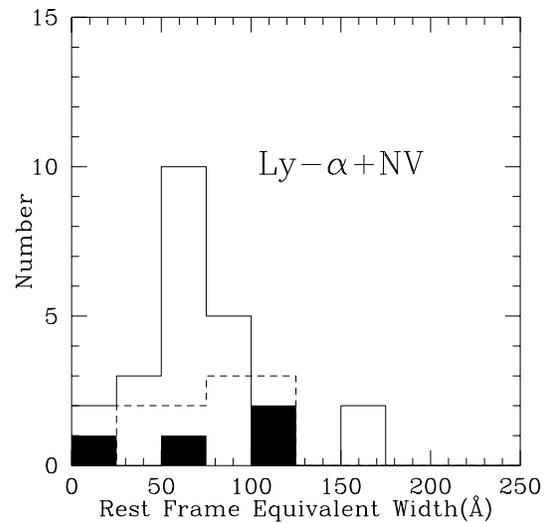,height=2.8in}}
\caption{Rest frame Ly-$\alpha$+NV equivalent width distribution
for $z>4$ quasars. The solid line is for 23 colour selected APM BRI
quasars from Storrie-Lombardi et al. (1996). The dotted line is for
the 10 grism-selected quasars from Schneider, Schmidt \& Gunn (1991a).
The solid bins are for radio selected $z>4$ quasars including both
those presented here and GB1508+5714 and PKS1251$-$407 (Shaver et al
1996a).}
\label{fig_ew}
\end{figure}

\section{Discussion}

GB1428+4217 is the most distant radio-loud object known and the 3rd
highest redshift quasar known (behind PC1247+3406 at $z=4.897$ and
PC1158+4635 at $z=4.73$, Schneider, Schmidt \& Gunn 1991b, 1989). 

GB1428+4217 has a tentative detection in the ROSAT All Sky Survey
(Brinkman et al. 1997) and an archival PSPC pointing (WGACAT, White
Giommi \& Angelini, 1994).  The extremely high apparent X-ray
luminosity of this object and the possibility that the object is
beamed are discussed in another Letter (Fabian et al 1997).  In Figure
4 we compare the equivalent widths of Ly-$\alpha$+NV for the four
known $z>4$ radio-selected quasars with other samples of $z>4$
quasars.  GB1428+4217 is notable in that it is amongst the lowest
equivalent width objects. This is consistent with a scenario in which
the continuum region is beamed. In addition, in contrast to the other
three radio loud $z>4$ quasars, the emission lines are relatively
broad.

After our INT and WHT observations 2 of the 38 POSS-APM blank fields
are quasars with $z>4$ and 23 are ruled out as being at $z>4$ quasars
(either based on spectroscopic observations, or deduced from their
blue $\rm B-R$ colours or extended appearance on the CCD images). 10
are still unidentified and a further 3 of the CCD identifications
still need conclusive spectra (1 of these has an inconclusive WHT
spectrum and 2 were not attempted).  Clearly, it is important to
determine the redshift status of these remaining 13 sources. Thus in
an effective sub-area of 0.73sr containing 302 sources, 97\% (all but
10) are optically identified and 96\% (all but 13) can be classified
as being a $z>4$ quasar or otherwise.

Therefore we can derive a lower limit to the space density at $z>4$
based on our sample, independently of any assumptions about the form
of evolution. In an area of 0.73sr, two objects with $z>4$ were found
among the CCD identifications. In our previous work we have found one
$z>4$ quasar among POSS-I identifications in an area of 3.66sr (Hook
et al 1995) which scales to 0.2 objects in 0.73sr. Thus a total of
$2.2\pm 1.4$ objects were effectively discovered in an area of 0.73sr
(where the uncertainties are from Poisson statistics).

At $z=5$ the flux limit of our sample, $\rm S_{5GHz}=0.2Jy$,
corresponds to a limiting radio power of $\rm 5.8
\times 10^{26} W Hz^{-1} sr^{-1}$, assuming a radio spectral index
$\alpha=0.0$.  For objects brighter than this limit we derive a space
density between $z=4$ and $z=5$ of $\rm 1.4 \pm 0.9\times 10^{-10}
Mpc^{-3}$. If any of the 13 objects above are found to lie at $z>4$,
this space density estimate would be increased, hence it should be
considered a lower limit. 
  
This can be compared to the value at $z=2$ from Dunlop \& Peacock
(1990) for flat-spectrum sources of the same luminosity. By
integrating their models $1-5$ over radio powers greater than $\rm 5.8
\times 10^{26} W Hz^{-1} sr^{-1}$, we derive
$\rm 2.9 \pm 0.2 \times 10^{-10} Mpc^{-3}$.  The study of Shaver et al (1996b)
deals with somewhat stronger sources ($\rm P\ge 1.1\times 10^{27}
WHz^{-1} sr^{-1}$) and since they show space densities normalised to
$z\sim 3$, is not possible to compare results directly.

\section*{Acknowledgments}
RGM thanks the Royal Society for support and IMH acknowledges a NATO
postdoctoral fellowship.  We thank Dan Stern for help with the
spectroscopic observations at Lick Observatory, Chris Benn for
carrying out the PATT observations on 1996 July 16th and David
Sprayberry for carrying out the PATT service program observations on
1996 July 19th. We also thank Rene Vermeulen for communicating
redshifts prior to publication.

\end{document}